\documentstyle[12pt]{article}

\begin{document}
\title{Cosmic String in Scalar-Tensor Gravity}
\author{M. E. X. Guimar\~aes \thanks{e-mail: marg@ccr.jussieu.fr} \\
\mbox{\small {Laboratoire de Gravitation et Cosmologie Relativistes}} \\
\mbox{\small {Universit\'e Pierre et Marie Curie - CNRS/URA 769}} \\
\mbox {\small {Tour 22/12, 4\`eme \'etage, B.C. 142}} \\
\mbox{\small {4, Place Jussieu 75252 Paris cedex 05, FRANCE}}}
\maketitle
\begin{abstract}
The gravitational properties of a local cosmic string in the framework of
scalar-tensor gravity are examined.
We find the metric in the weak-field approximation
and we show that, contrary to the
General Relativity case, the cosmic string in scalar-tensor gravitation
exerces a force on non-relativistic, neutral test particle.
This force is proportional
to the derivative of the conformal factor $A(\phi)$
and it is always attractive. Moreover, this force could have played an
important role at the Early Universe, although nowadays it can be
neglegible. It is also shown that the angular separation
$\delta\varphi$ remains unaltered for scalar-tensor cosmic strings.

\end{abstract}
\section{Introdution}

The scalar-tensor theories of gravity proposed by Bergmann \cite{st},
Wagoner \cite{wa} and Nordtverdt \cite{no} - generalizing the original
Brans-Dicke \cite{bd} theory - have been considerably revived
in the last years.
Indeed, the existence of a scalar field as a spin-0 component of the
gravitational interaction seems to be a quite natural prediction of
unification models such as supergravity or superstrings \cite{gr}.
Appart from the fact that scalar-tensor theories may provide a solution for
the problem of terminating inflation \cite{la,ga}, these theories by
themselves have direct implications for cosmology and for experimental
tests of the gravitational interaction: One expects that in the
Early Universe the coupling to matter of the
scalar component of the gravitational interaction would be of the same
order of the coupling to matter of
the long-range tensor component although
in the present time the observable total coupling strength of scalars
$(\alpha^2)$ is generically small \cite{da}. Besides, any gravitational
phenomena will be affected by the variation of the gravitational ``constant"
$G_{eff} \sim \tilde{\phi}^{-1}$. So, it seems worthwhile to analyse
the behaviour of matters in the framework of scalar-tensor theories,
specially those which originated in the Early Universe such as
topological defects, for exemple.

The aim of this paper is to study the modifications of the metric of a
local cosmic string in the framework of the scalar-tensor gravity. These
modifications are induced by the coupling of a scalar field to the
tensor field in the gravitational Lagragean. For simplicity, we will consider
a class of scalar-tensor theories where the potential $V(\tilde \phi)$
(or in Wagoner's notation \cite{wa}, $\lambda(\tilde\phi)$) is vanishing.
\newpage
The action describing these theories is (in Jordan-Fierz frame)

\begin{equation}
{\cal{S}} = \frac{1}{16\pi}\int d^4 x \sqrt{\tilde{g}} \;\;
[ \tilde{\phi}\tilde{R}
- \frac{\omega(\tilde{\phi})}{\tilde{\phi}} {\tilde{g}}^{\mu\nu}
\partial_{\mu}\tilde{\phi}\partial_{\nu}\tilde{\phi} ] +
{\cal{S}}_m[\Psi_m , {\tilde{g}}_{\mu\nu}] ,
\end{equation}
where ${\tilde{g}}_{\mu\nu}$ is the physical metric in this frame, $\tilde R$
is the curvature scalar associated to it and ${\cal{S}}_m$ denotes the action
of the general matter fields $\Psi_m$.
These theories are metric, which means that
matter couples
minimally to ${\tilde{g}}_{\mu\nu}$ and not to $\tilde{\phi}$.
For many reasons it is more convenient to work in the so-called
Einstein (conformal) frame, in which the kinetic terms of tensor and
scalar fields do not mix

\begin{equation}
{\cal{S}} = \frac{1}{16\pi G} \int d^4 x \sqrt{g} \;\;
[ R - 2 g^{\mu\nu}\partial_{\mu}\phi \partial_{\nu}\phi ] +
{\cal{S}}_m[\Psi_m , A^2(\phi)g_{\mu\nu}]  ,
\end{equation}
where $g_{\mu\nu}$ is the (unphysical) metric tensor in Einstein frame,
$R$ is
the curvature scalar associated to it and $A(\phi)$ is an arbitrary function
of the scalar field. Action (2) is obtained from (1) by a
conformal transformation in the physical metric

\begin{equation}
\label{co}
{\tilde{g}}_{\mu\nu} = A^2(\phi)g_{\mu\nu}  ,
\end{equation}
and by a redefinition of the quantities

\[
G A^2(\phi) = {\tilde \phi}^{-1} \;\; ,
\]
\[
\alpha^{2}(\phi) \equiv (\frac{\partial \ln A(\phi)}{\partial \phi})^2 =
[2\omega(\tilde \phi) + 3]^{-1} .
\]
It is important to remark that $\alpha(\phi)$ is the field-dependent coupling
strength between matter and scalar fields. In the particular
case of Brans-Dicke theory, $A(\phi)$ has the following dependence on $\phi :
A(\phi) = exp[2\alpha\phi]$, with $\alpha(\phi)=\alpha=$const.
In the Einstein frame, the field equations are

\begin{eqnarray}
R_{\mu\nu} & = & 2\partial_{\mu}\phi\partial_{\nu}\phi + 8\pi G(T_{\mu\nu}-
\frac{1}{2}g_{\mu\nu} T),  \nonumber \\
& & \Box_{g}\phi = -4\pi G \alpha(\phi) T .
\end{eqnarray}
The first of the above equations can also be written in terms of
the Einstein tensor $G_{\mu\nu}$

\[
G_{\mu\nu} = 2 \partial_{\mu}\phi\partial_{\nu}\phi -
g_{\mu\nu}g^{\alpha\beta}\partial_{\alpha}\phi\partial_{\beta}\phi
+ 8\pi GT_{\mu\nu} .
\]
The energy-momentum tensor is defined as usual

\[
T_{\mu\nu} \equiv \frac{2}{\sqrt{g}}
\frac{\delta {\cal{S}}_m[A^{2}(\phi)g_{\mu\nu}]}{\delta g_{\mu\nu}} ,
\]
and in the Einstein frame it is no longer conserved

\begin{equation}
\nabla_{\nu} T^{\nu}_{\mu} = \alpha(\phi)T \nabla_{\mu}\phi .
\end{equation}
It is clear from transformation (\ref{co}), that we can relate
quantities from both frames such that $\tilde{T}^{\mu\nu} =
A^{-6} T^{\mu\nu}$  and
$ \tilde{T}^{\mu}_{\nu} = A^{-4} T^{\mu}_{\nu}$ .

In what follows, we will search for a regular solution of an
isolated static straight
cosmic string in the scalar-tensor gravity described above. Hence,
the cosmic string arises from the action of the Abelian-Higgs model where
a charged scalar Higgs field $\Phi$ minimally couples to the $U(1)$ gauge
field $A_{\mu}$

\begin{equation}
\label{cs}
{\cal S}_{m} = \int dx^4 \sqrt{\tilde g} \;\; [\frac{1}{2} D_{\mu}\Phi
D^{\mu}\Phi^* - \frac{1}{4}F_{\mu\nu}F^{\mu\nu} - V(\mid \Phi \mid) ] ,
\end{equation}
with $D_{\mu} \equiv \partial_{\mu} + ieA_{\mu}$, $F_{\mu\nu} \equiv
\partial_{\mu}A_{\nu} -\partial_{\nu}A_{\mu}$ and the Higgs potential
\newline
$V(\mid\Phi\mid) = \lambda (\mid\Phi\mid^2 -\eta^2)^2$ . $e, \lambda$ and
$\eta$ are positive constants, $\eta$ being the characteristic energy scale
of the symmetry breaking (for typical grand unified theories (GUT),
$\eta \sim 10^{16} GeV $).

We confine our attention to the static configurations of vortex type about
the $z$-axis. In cylindrical coordinate system $(t,z,\rho ,\varphi)$ with
$\rho \geq 0$ and $0 \leq \varphi < 2\pi$, we impose the following form for
the scalar $\Phi$ and the gauge $A_{\mu}$ fields \footnote{For simplicity,
we set the winding number $n=1$.}

\begin{equation}
\Phi \equiv  R(\rho)e^{i\varphi}  \;\;\;\; and \;\;\;\;
A_{\mu} \equiv \frac{1}{e}[P(\rho) - 1]\delta^{\varphi}_{\mu}
\end{equation}
where $R,P$ are functions of $\rho$ only. Moreover, we require that
these functions are regular and finite everywhere and that they satisfy the
usual boundary conditions for vortex solutions \cite{ol}

\[
R(0) = 0  \;\;\;\; and \;\;\;\; P(0)=1 \]
\begin{equation}
\lim_{\rho\rightarrow \infty}R(\rho) = \eta \;\;\;\;  and \;\;\;\;
\lim_{\rho\rightarrow \infty}P(\rho)=0 .
\end{equation}

In General Relativity, a metric for the cosmic string described by the
action (6) above has been already found in the
assymptotic limit by Garfinkle \cite{gar} and exactly by Linet \cite{li1}
provided the particular relation $e^2=8\lambda^2$ between the constants $e,
\lambda$ is satisfied. The question is whether one can find a solution for
the cosmic string in the context of the scalar-tensor gravity.
We anticipate that an exact solution for both the matter fields and Einstein
equations is impossible to be determined analytically. So that, we will
consider the weak-field approximation for this solution in the same way as
Vilenkin did in the framework of General Relativity \cite{vi}. In fact, the
weak-field approximation breaks down at large distances from the cosmic
string. Therefore,
we assume that, at large distances, the $\phi$ dependence on the
right-hand side of the first of Einstein eqs. (4) must dominate over the
$T^{\mu}_{\nu}$ term \footnote{As shown by Laguna-Castillo and Matzner
\cite{la2}, the $T^{\mu}_{\nu}$ components vanish far from the cosmic string
in General Relativity. Gundlach and Ortiz \cite{gu} showed that the same
occurs if one considers cosmic string in Brans-Dicke theory. One must expect
that this result remains valid in general scalar-tensor theories.}. So that,
we can neglect the energy-momentum tensor in Einstein eqs. (4) and find the
vacuum solution as an asymptotic behaviour of $g_{\mu\nu}$ and $\phi$ far
from the cosmic string and, then, match it with the metric in the weak-field
approximation.

The plan of this work is as follows. In section 2, we find the exact metric
for the vacuum Einstein equations. In section 3, we find the metric
of the cosmic string in the
weak-field approximation of the scalar-tensor gravity and analyse under which
conditions it can be matched to the vacuum metric of section 2.
In passing, we show that, contrary to the General Relativity case \cite{vi},
the cosmic string in scalar-tensor
gravity exerces a force on a test-particle. This force is
always attractive and proportional to $A'(\phi) \sim \alpha(\phi)$.
However, the angular separation $\delta\varphi$ remains
unaltered ($\delta\varphi\sim 10^{-5} rad$) for scalar-tensor cosmic strings.
Finally, in section 4 we add some concluding remarks.

\section{The Vacuum Metric in Scalar-Tensor \newline Gravity}

In this section we will show that there exists an exact static vacuum
metric which is solution to the Einstein equations in scalar-tensor theories.
Since this solution is supposed to match the (weak-field) solution of the
cosmic string, it seems natural to impose that
the vacuum spacetime has the same symmetries than the string.
So that, we write the following metric

\begin{equation}
\label{met}
ds^2=g_1(\rho) dt^2 - g_2(\rho) dz^2 -d\rho^2 -g_3(\rho) d\varphi^2 ,
\end{equation}
where $g_1, g_2, g_3$ are functions of $\rho$ only, and $(t,z,\rho ,\varphi)$
are cylindrical coordinates with $\rho \geq 0$ and $0 \leq \varphi < 2\pi$.
Defining $u \equiv (g_1 g_2 g_3)^{1/2}$,
the Einstein equations (4) in vacuum are

\begin{equation}
R^{i}_{i} = \frac{1}{2u}[u\frac{g'_i}{g_i}] =0 , \;\;\;\;  (i = t,z,\varphi)
\end{equation}
\begin{equation}
G^{\rho}_{\rho} = -\frac{1}{4} [ \frac{g'_1 g'_2}{g_1 g_2} + \frac{g'_1 g'_3}
{g_1 g_3} + \frac{g'_2 g'_3}{g_2 g_3} ] = - (\phi')^2 ,
\end{equation}
\begin{equation}
\frac{1}{u} \frac{d}{d\rho} (u \phi') = 0 ,
\end{equation}
where the prime means derivative with respect to $\rho$.
Moreover, since $T^{\mu}_{\nu} =0$, the following expression is also valid

\begin{equation}
\label{soma}
\Sigma_{i} R^i_i = \frac{u''}{u} = 0  \;\;\;\;  (i= t,z,\varphi) .
\end{equation}
{}From (\ref{soma}), it follows that $u$ is a linear function of $\rho$
($u \sim B\rho$). This result enables us to solve eqs. (10) and (12),
respectively

\begin{equation}
g_i = k_i^{(0)}(\frac{\rho}{\rho_0})^{k_i}  \;\;\;\; and,
\end{equation}
\begin{equation}
\phi = \phi_0 + \kappa \ln \rho/\rho_0 ,
\end{equation}
where $B, k_i^{(0)}, k_i$ and $\kappa$ are constants to be
determined later. Combining the solution for $u$ and (14) and (15)
in the differential eq. (11), we find
the following relations between the constants

\[
(k_1^{(0)}k_2^{(0)}k_3^{(0)})^{1/2}= B ,
\]
\[
k_1 + k_2 + k_3 = 2 \;\;\;\; and,
\]
\[
k_1k_2 + k_1k_3 + k_2k_3 = 4\kappa^2 .
\]
Moreover, if we suppose that the Lorentz invariance along the $z$-axis is
still valid ($g_1 = g_2)$, we finally get

\begin{equation}
k_1^{(0)}(k^{(0)}_3)^{1/2} = B ,
\end{equation}
\begin{equation}
k_3 = 2 - 2k_1  \;\;\;\; and,
\end{equation}
\begin{equation}
\kappa^2 = k_1 (1- \frac{3}{4}k_1) .
\end{equation}
Since the constant $k_1^{(0)}$ can always be absorbed by a
redefinition of $t$ and $z$, we get the final form for
the vacuum metric

\begin{equation}
ds^2 = (\frac{\rho}{\rho_0})^{k_1}(dt^2 - dz^2)-d\rho^2-(\frac{\rho}{\rho_0})
^{2 - 2k_1}B^2 d\varphi^2 ,
\end{equation}
in which $k_1$ (and consequently $k_3$), $B$ and $\kappa$ will be fully
determined after the introduction of matter fields.
It is worthwhile to mention that in the particular case
of Brans-Dicke theory (i.e., substituting $A^2(\phi) = exp[2\alpha\phi]$
in expression (19)),
we get the same result as Gundlach and Ortiz in ref. \cite{gu} if we correct
the expression $(r-r_0)$ to $r/r_0$ in metric (14) in their paper.
Note that (19) can also be written in Taub-Kasner form \cite{fil}
after a suitable transformation of coordinates.

The Ricci tensor in the vacuum spacetime described by (19) is regular
(vanishes everywhere) if and only if $\phi=\phi_0=$constant
(i.e., $\kappa =0$). This should not be surprising in views of the
structure of Einstein eqs. (10-12), in particular eq. (11). Moreover,
$\kappa =0$ implies that the only allowed values for $k_1$ are $k_1=4/3$
and $k_1=0$. This latter value corresponds to the conical
metric and only
in this case $B^2= k_3^{(0)}$ can be interpreted as the deficit angle.
As we will see in the next section, these values ($k_1=0$ and $k_1=4/3$)
are precisely the needed conditions for matching metric (19) to the
metric of the cosmic string in the weak-field approximation.

Finally, let us remind that the {\em physical} vacuum metric is obtained
by multiplying (19) by the conformal factor $A^2 (\phi)$, with $
\phi = \phi_0 + \kappa\ln \rho/\rho_0$
and $\kappa^{2} = k_1 (1- \frac{3}{4}k_1)$.

\section{Cosmic String Solution in Scalar-Tensor \newline
Gravity: the weak-field approximation}

The full Einstein equations are

\newpage

\begin{equation}
R^t_t  =  R^z_z = \frac{1}{2u}[u \frac{g'_1}{g_1}]'=8\pi G(T^t_t -\frac{1}
{2}T) ,
\end{equation}
\begin{eqnarray}
R^{\rho}_{\rho} & = & \frac{1}{2}[2 (\frac{g''_1}{g_1}) -
(\frac{g'_1}{g_1})^2
+ (\frac{g''_3}{g_3}) - \frac{1}{2}(\frac{g'_3}{g_3})^2] \nonumber \\
& & = - 2(\phi')^2 + 8\pi G(T^{\rho}_{\rho}-\frac{1}{2}T) ,
\end{eqnarray}
\begin{equation}
R^{\varphi}_{\varphi} = \frac{1}{2u}[u\frac{g'_3}{g_3}]' = 8\pi G
(T^{\varphi}_{\varphi} -\frac{1}{2}T) ,
\end{equation}
\begin{equation}
\frac{1}{u}[u\phi']' = 4\pi G\alpha(\phi) T ,
\end{equation}
with the matter fields equations (where form (7) for the material fields were
imposed)

\begin{equation}
R'' + R' \left( \frac{g'_1}{g_1} +\frac{1}{2}\frac{g'_3}{g_3} - 2\alpha(\phi)
\phi' \right) -R (g_3^{-1}P^2 +4\lambda A^2 (R^2-\eta^2))=0 ,
\end{equation}
\begin{equation}
P''+P'\left(\frac{g'_1}{g_1}+\frac{1}{2}\frac{g'_3}{g_3}-4\alpha(\phi)\phi'
\right) -e^2A^2R^2P^2=0 .
\end{equation}
The non-vanishing components of $T^{\mu}_{\nu}$ are

\begin{eqnarray}
& T^t_t & = T^z_z =  \frac{1}{2}A^2(\phi) \left [(R')^2+g_3^{-1}R^2P^2 +
\frac{1}{e^2}g_3^{-1}A^{-2}(P')^2 + 2\lambda A^2(R^2-\eta^2)^2\right],
\nonumber \\
& & T^{\rho}_{\rho} = -\frac{1}{2}A^{2}(\phi) \left[(R')^2-g_3^{-1}R^2P^2+
\frac{1}{e^2}g_3^{-1}A^{-2}(P')^2-2\lambda A^2(R^2-\eta^2)^2\right],
\nonumber \\
& & T^{\varphi}_{\varphi} = \frac{1}{2}A^{2}(\phi) \left [(R')^2 -g_3^{-1}
R^2P^2-\frac{1}{e^2}g_3^{-1}A^{-2}(P')^2 +2\lambda A^2(R^2-\eta^2)^2 \right]
\end{eqnarray}

In views of the impossibility to find an exact solution for eqs. (20-25),
we will consider the metric of a cosmic string in scalar-tensor gravity
in the weak-field approximation. In fact, this approximation can be
justified only
if we consider the scalar field $\phi$ as a small perturbation on the
gravitational field of the cosmic string. So that, it may be expanded in
terms of a small parameter $ \epsilon$ about the values $\phi =\phi_0$ and
$g_{\mu\nu}=\eta_{\mu\nu}$:

\[
\phi = \phi_0 + \epsilon \phi_{(1)} + \ldots
\]
\[
g_{\mu\nu} = \eta_{\mu\nu} + \epsilon h_{\mu\nu} +\ldots
\]
\[
A(\phi) = A(\phi_0) + \epsilon A'(\phi_0)\phi_{(1)} +\ldots
\]
\[
T^{\mu}_{\nu} = T_{(0)\nu}^{\mu} + \epsilon T^{\mu}_{(1)\nu} +\ldots
\]
The term $(\phi')^2$ is neglected in the process of linearisation of the
Einstein eqs. (20-23). Moreover, in this approximation,
the $T^{\mu}_{(0)\nu}$
term is obtained from (26) by a limit process $\lambda \rightarrow \infty$
\cite{li1}. Therefore, it tends to the Dirac distribution on the hypersurface
$t=$ and $z$=constant. In this way, the linearized
equations reduce to those of General Relativity \cite{vi} (except that in
our case $T^{\mu}_{(0)\nu}$ and $h_{\mu\nu}$ carry the conformal factor
$A^{2}(\phi)$)

\begin{equation}
\nabla^2 h_{\mu\nu} = 16\pi G (T^{\mu}_{(0)\nu} -
\frac{1}{2}\eta_{\mu\nu}T_{(0)}).
\end{equation}
Besides, the linearized equation for the scalar field is
\begin{equation}
\nabla^2\phi_{(1)} = 4\pi G \alpha(\phi_0)T_{(0)} .
\end{equation}
The solution for eq. (28) is

\begin{equation}
\phi_{(1)} = 4 \mu G  A^2(\phi_0) \alpha(\phi_0)\ln (\rho/\rho_0).
\end{equation}
This solution matchs with the vacuum metric if and only if

\begin{equation}
\kappa_{lin} = 4G\mu \alpha(\phi_0) A^2(\phi_0) .
\end{equation}
{}From (18) we see that
the only allowed values for $k_1$ are $k_1 = 0 + O(G^2\mu^2)$ and
$k_1 = 4/3 + O(G^2\mu^2)$. The only physical-meaning result is $k_1 = 0$
plus neglegible corrections of order $O(G^2\mu^2)$;
the other value corresponding to nonphysical metric.

The solutions for eq. (27) differs from those found by Vilenkin \cite{vi}
by a conformal factor $A^{2}(\phi)$. However, the procedure is the same as
in his paper. We obtain

\begin{eqnarray}
ds^2 & = & A^2(\phi_0) (1+ 8\mu GA^{2}(\phi_0)\alpha^{2}(\phi_0)
\ln(\rho/\rho_0)) \{ dt^2 -dz^2 -d\rho^2
\nonumber \\
& & - [1- 8\mu G A^2(\phi_0)](\rho/\rho_0)^2 d\varphi^2 \} .
\end{eqnarray}
Therefore, metric (31) represents an isolated
scalar-tensor cosmic string in the weak-field approximation.
It is important to remark that the cosmic string in this theory
exerces a force on a neutral test-particle given by

\begin{eqnarray}
f^{\rho} & = & -\frac{1}{2}\frac{d h_{00}}{d\rho} \nonumber \\
& & = -4 \mu GA^{2}(\phi_0)\alpha^2(\phi_0)\frac{1}{\rho}
\end{eqnarray}
and it is always attractive. Moreover, although this force seems to be
neglegible nowadays, in the Early Universe it was not so in views of the
factor $G A^{2}(\phi)$ in (32). This result differs from Brans-Dicke
\cite{gu,ba} because the parameters in this theory are constants.
The light deflection $\delta\varphi$ is $\delta
\varphi=4\pi GA^2(\phi_0)=4\pi{\tilde G}_0 \mu$, where ${\tilde G}_0$ is the
effective Newtonian constant. Therefore, the value $\delta\varphi \sim
10^{-5}$rad for GUT strings remains valid in scalar-tensor gravity as well
as in General Relativity. It means that if one believes that the effect of
double image of some astrophysical objects is due
to the presence of a cosmic
string between them and the observer, the fact that the angular separation
$\delta\varphi \sim 10^{-5} rad $ remains unaltered reveals that it would be
impossible to distinguished, at least by this observation, scalar-tensor
cosmic strings from General Relativity ones.

\section{Conclusion}

In this work we found the metric of a cosmic string in the weak field
approximation of scalar-tensor gravity.
It is shown that this metric is conformal to the metric found by
Vilenkin in the framework of General Relativity. We showed that,
in contrast with the General Relativity case, the cosmic string in
scalar-tensor gravity exerces a force on a test particle and that this force
is always attractive. However, the light deflection remains the same, at
least in the linear approximation. It is worthwhile to point out that in
this paper we generalized the works made by Gundlach and Ortiz \cite{gu} and
recently by Barros and Romero \cite{ba}, both in
the framework of Brans-Dicke theory of gravity.

\section*{Acknowledges}

The author gratefully acknowledges Profs. Bernard Linet and
Pierre Teyssandier for unvailable comments, suggestions
and critical reading of this manuscript. This work is supported by
a grant from CNPq (Brazilian government agency).

\end{document}